# Unstable Retention Behavior in MIFIS FEFET: Accurate Analysis of the Origin by Absolute Polarization Measurement

Song-Hyeon Kuk, Kyul Ko, Bong Ho Kim, Jae-Hoon Han, *Member*, Sang-Hyeon Kim,

**Abstract**— Ferroelectric field-effect-transistor (FEFET) has emerged as a scalable solution for 3D NAND and embedded flash (eFlash), with recent progress in achieving large memory window (MW) using metal-insulator-ferroelectric-insulator-semiconductor (MIFIS) gate stacks. Although the physical origin of the large MW in the MIFIS stack has already been discussed, its retention characteristics have not been explored yet. Here, we demonstrate MIFIS FEFET with a maximum MW of 9.7 V, and show that MIFIS FEFET has unstable retention characteristics, especially after erase. We discover the origin of the unstable retention characteristics and prove our hypothesis with absolute polarization measurement and different operation modes, showing that the unstable retention characteristics is a fundamental issue. Based on the understanding, we discuss a novel charge compensation model and promising engineering methodologies to achieve stable retention in MIFIS FEFET.

## I.  INTRODUCTION

Word-line (W/L) stacking in charge-trap-flash (CTF)-based 3D NAND recently poses challenges with longer strings, resulting in a need for FEFET-based NAND (FE-NAND) for further scaling (Fig. 1(a)) [1-3]. Also, ferroelectrics-based eFlash has been proposed [4-5], for enhancing the speed and scalability of the conventional eFlash such as two-transistor NOR (2T-NOR) and split-gate flash devices. The novel trend of the research is based on the excellent properties of the ferroelectric doped $HfO_2$ such as scalability, complementary-metal-oxide-semiconductor (CMOS) compatibility and thermal stability.

However, the small MW has been considered a critical bottleneck for FEFET as an emerging memory. Preceding studies have revealed that screening charges, which are mostly trapped near the interfaces of the ferroelectric layer and the channel, result in narrowing MWs [6]. A study even showed that only 10% of the remnant polarization ($P_r$) can change the channel potential [7].

To overcome the bottleneck, recent progress focused on gate stack engineering. Notably, a MW of 10.5 V was achieved by using an MIFIS stack, satisfying the requirements of 3D NAND (e. g. < 20 nm gate stack thickness) [8]. Employing FE-NAND was expected to reduce the total cell height by up to −44 % in an advanced product generation [9]. The large MW in MIFIS FEFET was attributed to "charge compensation" of polarization by trapped charges at the top interlayer (top IL), as Fig. 1 [10-12].

However, most prior studies only present the maximum MW, neglecting crucial reliability data such as retention characteristics [8-12]. Data retention is pivotal in determining the feasibility and practicality of memory devices, necessitating detailed discussion.

Additionally, while previous studies have reported charge compensation models in MIFIS FEFETs, all of them predominantly rely on simulation data of polarization switching to support their claims. The absence of the experimental data is because the direct observation of polarization switching in the FEFET device requires meticulous and comprehensive measurement processes [6-7]. Only a few groups have used a methodology, large-signal pulsed quasi-static split *C-V* (QSCV), to directly quantify polarization switching and to experimentally validate their models of MFIS FEFETs [6-7, 13-14].

In this study, we demonstrate MIFIS FEFET with a maximum MW of 9.7V and explore the retention characteristics. Because it turns out that MIFIS FEFET has unstable retention characteristics, we discuss the physical origin of the instability by 1) measuring QSCV, and retention of threshold voltage ($V_{th}$) and remnant polarization ($P_r$); 2) Comparison between hole injection and electron de-trapping modes. Based on our findings, we provide promising engineering approaches to achieve stable retention in MIFIS FEFET for flash memory applications. The methodology used in this work does not need any device simulation to support the proposed charge compensation model. Therefore, our study provides comprehensive experimental data of absolute polarization and absolute $P_r$ retention for the first time which have not been provided by any other literature.

Manuscript received MM, DD, YYYY. This work was supported partly by NRF of Korea grant (No. RS-2023-00215860, 2022M3I8A107725712, 2022R1C1C1007333), partly by BrainKorea21 FOUR, partly by IC Design Education Center (IDEC), and partly by Korea Institute of Science and Technology (KIST) Institutional Program (2E32942).
Song-Hyeon Kuk, Bong Ho Kim and Sang-Hyeon Kim* are with School of Electrical Engineering, Korea Advanced Institute of Science and Technology (KAIST), Daejeon 34141, Republic of Korea (*e-mail: shkim.ee@kaist.ac.kr).
Kyul Ko and Jae-Hoon Han* are with Korea Institute of Science and Technology (KIST), Seoul 02792, Republic of Korea (*e-mail: hanjh@kist.re.kr).

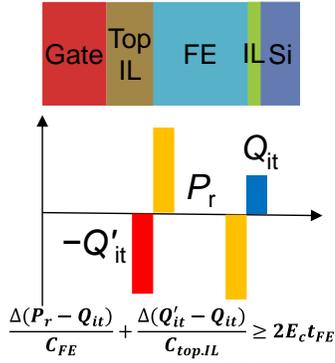

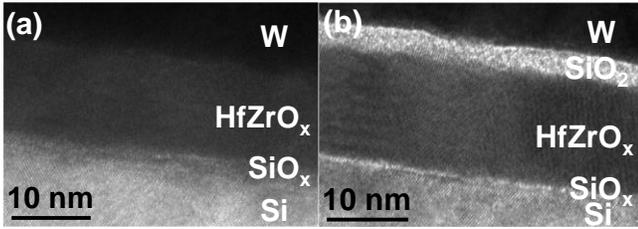

$$\frac{\Delta(P_r - Q_{it})}{C_{FE}} + \frac{\Delta(Q'_{it} - Q_{it})}{C_{top,IL}} \geq 2E_c t_{FE}$$

Fig. 1. A charge compensation model in MIFIS FEFET from the previous literature. Because most of $P_r$ are screened by trapped charges at the bottom IL ($Q_{it}$), MFIS FEFET exhibits small MWs (< 4 V). However, in MIFIS FEFET, trapped charges at the top IL ($Q'_{it}$) compensate $Q_{it}$ and finally, the effect of $P_r$ can be fully seen from the channel. Therefore, the MW can be enlarged, up to 10.5 V [8, 9].

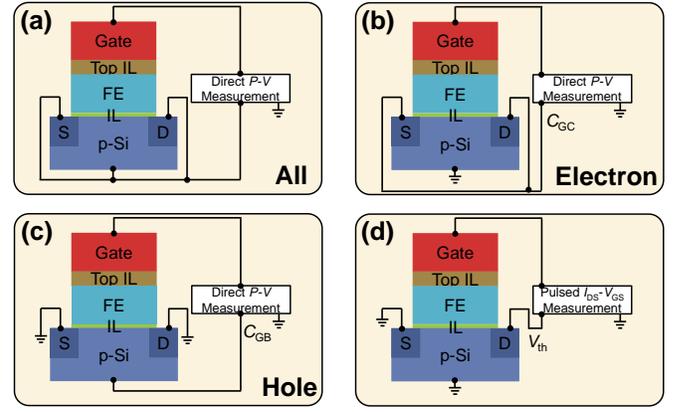

Fig. 3. Schematics of measurement set-ups. (a) QSCV measurement for the all carriers (electron, hole). (b) QSCV measurement for electron and (c) for hole. (d) Pulsed $I_{DS}$-$V_{GS}$ measurement. $V_{DS}$ of 0.2 V is used in this work.

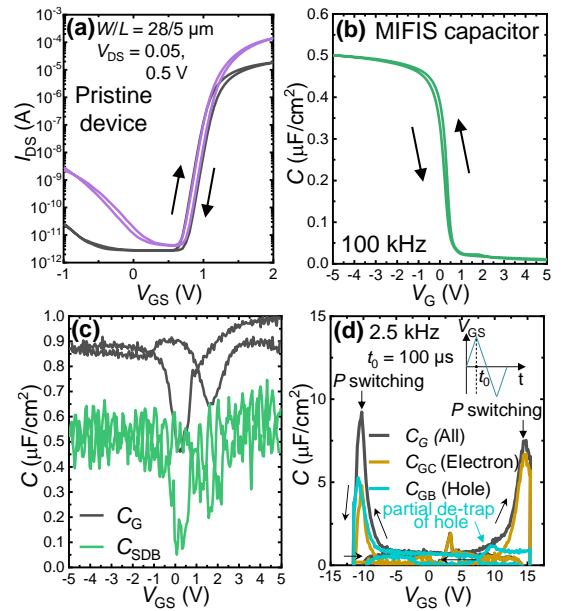

Fig. 4. (a) DC $I_D$-$V_G$ of the pristine MIFIS FEFET. Low $V_{GS}$ biasing ($|V_{GS}|$<5 V) did not switch ferroelectric polarization. (b) Measured small-signal $C$-$V$ curves at 100 kHz (AC voltage = 30 mV) (c) Large-signal pulsed $C$-$V$ curves at 2.5 kHz directly measured in the MIFIS FEFET device. $C_G$ is a capacitance measured at the gate electrode and $C_{SDB}$ is a capacitance measured from the source, the drain and the body [6]. It shows that the pristine device with low $V_{GS}$ biasing shows no ferroelectric polarization switching. $C_{SDB}$ data show noises due to current measurement limit, and $C_G$ and $C_{SDB}$ are different because $C_G$ includes transient capacitance charging at the measurement cable line. Therefore, $C_{SDB}$ is the accurate gate capacitance of the FEFET device. (d) QSCV of MIFIS FEFET at 2.5 kHz. Polarization switching is clearly observed, and the contributions of electrons and holes to polarization switching are presented.

Fig. 2. Cross-sectional TEM images of (a) MFIS (control) and (b) MIFIS FEFET gate stacks. The top IL (SiO2) has a thickness of 4 nm.

## II. ASYMMETRIC IMPACTS OF PROGRAM AND ERASE ON MEMORY WINDOW ENLARGEMENT

To fabricate MIFIS FEFETs on Si substrate, Si wafer was cleaned by SC-1 solution, and the source and the drain regions were doped highly by ion implantation (PH3 gas, 30 keV, 5×10^15 cm^-2, tilted angle: 7°) and dopant activation (RTA, 10 seconds at 1000 ºC). The wafer was cleaned by SC-1 and diluted HF several times before depositing gate oxides. HfZrO_x (HZO) was deposited (ZrO_x 12 cycles, HfO_x 13 cycles, laminated, 7 layers, 250 °C) by atomic layer deposition (ALD). The top IL (SiO2) was deposited by ALD (at 150 °C) and W gate and metal pad were formed. SiO2 thicknesses are optimized (from 1 nm to 5 nm) to achieve the large MW. Post-metallization-annealing (PMA) was carried out by rapid-thermal-annealing (RTA) at 430 ºC, and Al for S/D pads was formed, followed by additional RTA at 300 ºC for the contact. Control MFIS FEFET was also fabricated with the same process except the top IL deposition.

Figs. 2(a) and 2(b) show the cross-sectional transmission-electron-microscopy (TEM) images of the fabricated MFIS and MIFIS FEFETs, revealing the top IL thickness (4 nm), the bottom IL (SiO_x) thickness (0.7 nm) and the total gate stack thickness (17.8 nm). Here, a thick HZO film (13.8 nm) was inevitably used, because the HZO film has a high dielectric constant (28.8). Considering the Gauss law, the high dielectric constant leads to a small potential drop in the ferroelectric film, which is not enough to switch ferroelectric polarization in the MIFIS stack. Also, owing to the same reason, the thickness of top IL (SiO2) should be optimized to guarantee polarization switching, low gate leakage and reasonable amounts of trap sites.

Fig. 3 shows schematics of measurement set-ups used in this work. Figs. 3(a), 3(b) and 3(c) are used, to separately evaluate the contribution of polarization switching by electrons and holes during QSCV measurement. The pulsed $I_{DS}$-$V_{GS}$ set-up in Fig. 3(d) is used for measuring $V_{th}$ in the devices.

Fig. 4(a) shows DC $I_D$-$V_G$ curves in the pristine MIFIS FEFET device. A pristine device is tested with low gate voltages ($|V_{GS}| < 5$ V) not to stimulate ferroelectric polarization [15]. This method also has been carried out to examine conventional flash memory devices in preceding literature [16].

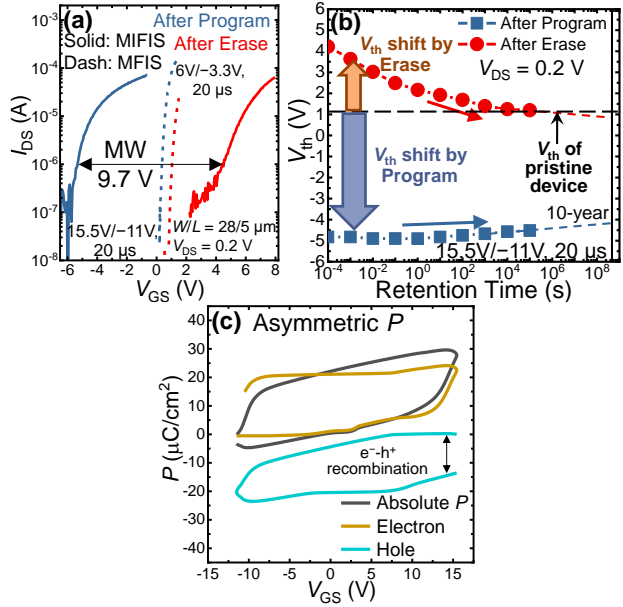

Fig. 5. (a) Obtained memory window by pulse measurement after program/erase in MFIS and MIFIS FEFETs. (b) Retention characteristics of $V_{th}$ in MFIS and MIFIS FEFETs. MW decreases −37% after $10^5$ s. 66% of the MW ($V_{th}$ shift) are made by program, and 34% are by erase, considering $V_{th}$ of the pristine device which did not switch polarization by low $V_{GS}$ biasing ($|V_{GS}|<5$ V) in Fig. 4(a). (c) Absolute polarization measurement (at 2.5 kHz) using the carrier dynamics. It shows that polarization switching is asymmetric, and erase pulses (negative pulses) do not switch polarization fully.

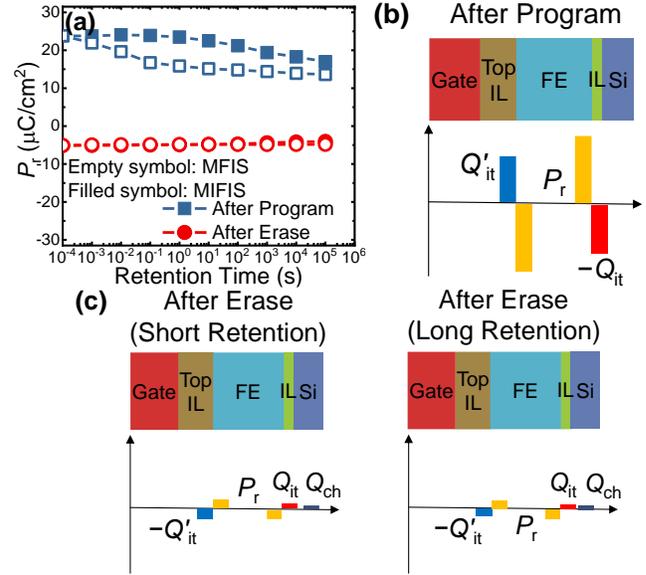

Fig. 6. (a) Retention characteristics of $P_r$ in MFIS and MIFIS FEFETs. Schematics of suggested charge compensation models of (b) after program and (c) after erase with short and long retention time. Asymmetric polarization switching leads to unstable retention behaviors after erase.

Narrow hysteresis by charge trapping and $V_{th}$ of 1.1 V (defined as the $V_{GS}$ at $I_{DS} = 1$ μA in this paper) are observed. A small-signal $C$-$V$ curve of an MIFIS capacitor at 100 kHz is shown in Fig. 4(b). It can be compared to large-signal pulsed QSCV curves in Fig. 4(c). The matching of small-signal capacitance and $C_{SDB}$ is the evidence of the low $V_{GS}$ operation without stimulating polarization [6]. On the other hand, when the high $V_{GS}$ pulses are applied as Fig. 4(d), clear polarization switching peaks are observed.

Furthermore, QSCV enables detailed analysis of the respective contributions of electrons and holes during program and erase [6]. Specifically, electron contribution dominates during program while partial hole de-trapping is also observed. On the other hand, hole becomes the primary factor of polarization switching during erase, while also partial electron de-trapping from the gate stack to the channel is observed.

Fig. 5(a) shows that the maximum MW of 9.7 V is achieved in MIFIS FEFET, while MFIS FEFET shows a MW of 0.9 V. However, it turns out that the maximum MW achieved at the short retention time (100 μs) is easily deteriorated as the retention time evolves. Fig. 5(b) shows that the MW diminishes fast, mostly by the erased state. In addition, considering the $V_{th}$ of the pristine device, it is obvious that the program operation shifts $V_{th}$ much more effectively than the erase operation. From the $V_{th}$ of the pristine device, program shifts 6.0 V (at a short retention time), while erase shifts only 3.1 V.

The significant asymmetry between the program and erase is further explored by measuring absolute polarization in MIFIS FEFET [15]. In the MFIS structure, it has been discussed that the ferroelectric polarization cannot be fully switched in certain cases due to the carrier dynamics. Especially, the MFIS FEFET devices with the Si channel typically show significantly asymmetric polarization switching behaviors by program/erase pulses [13]. Erase pulses (negative pulses) cannot switch polarization properly, while program pulses switch polarization fully. This points out that the significant asymmetric behaviors of programmed/erased states in MIFIS FEFET must be originated from the asymmetric polarization switching.

Hence, absolute polarization using large-signal pulsed quasi-static split $C$-$V$ (QSCV) is measured as in Fig. 5(c). The absolute polarization is different from the "centering" process because it uses carrier dynamics to obtain the absolute values of polarization. For instance, when the channel is fully inversed, holes must not be seen in the channel. On the other hand, when the channel is accumulated, the electron must not be observed at the channel. Using this physics, one can determine the hole carrier density during program ($V_{GS} > 13$ V) as zero. Also, the electron carrier density during erase ($V_{GS} < -9$ V) can be considered zero. Therefore, by plotting polarization curve measured by the set-up in Figs. 3(c), 3(b) and 3(a), one can get the "absolute" polarization value. This methodology is important for understanding the unstable retention characteristics of MIFIS FEFET because it would reveal that polarization switching is asymmetric after program/erase.

The obtained absolute polarization in Fig. 5(c) shows asymmetric curves, the same as that of MFIS FEFET [13]. Any previous literature on MIFIS FEFET has not carefully considered this. Asymmetric polarization switching by positive/negative pulses is due to the low dielectric constant of the bottom IL, SiO$_x$ [15, 17]. When the gate voltage is applied to the gate metal, it distributes across top IL, the FE layer, the bottom IL and the channel. According to Gauss' law, an oxide with a low dielectric constant receives more gate voltage than one with a high dielectric constant if the thicknesses are same. In this regard, it has been reported that a certain amount of initial charge trapping rather effectively induces polarization switching in MFIS FEFETs with Si channels [18-20]. It has been known that the electron takes a more appropriate role of trapping-assisted polarization switching than the hole does [20-23]. This is why the polarization curve is asymmetric in both MFIS and MIFIS FEFETs, and erase pulses do not flip polarization properly.

To summarize, the strong ferroelectric switching after the program operation enlarges the memory window, whereas the weak ferroelectric switching after erase shifts the $V_{th}$ only slightly. However, as we discussed above, the increase of MW in MIFIS FEFET is attributed to the charge compensation at the top IL. To further examine the trapped charges at the top IL and their relations with asymmetric polarization switching, we also discuss the $P_r$ retention.

Fig. 6(a) shows the $P_r$ retention measured in MFIS and MIFIS FEFET by QSCV, considering absolute polarization. Considering the de-polarization field, MIFIS FEFET should show the degraded $P_r$ retention compared to MFIS FEFET, in the ideal case (without charge trapping). However, in reality, trapped charges at the top IL screen and compensate the polarized domains in the ferroelectric film, which even enhances the $P_r$ retention in MIFIS FEFET compared to MFIS FEFET. Especially, the retention of $P_r$ after program is notably enhanced, which indicates that trapped holes at the top IL after program take critical roles of stabilizing $P_r$ and shifting $V_{th}$ by charge compensation.

Thus, considering the absolute polarization curve, $V_{th}$ and $P_r$ retention, the charge compensation model can be described as Figs. 6(b) and 6(c). Note that the amount of flipped polarization in Figs. 6(b) and 6(c) are different—this is due to asymmetric polarization switching found in this work. After program, $P_r$ is large enough (~$10^{14}$ cm$^{-2}$), which induces significant hole trapping at the top IL ($Q'_{it}$) to screen $P_r$. This compensates the trapped electrons at the bottom IL ($Q_{it}$) and effectively shifts $V_{th}$. On the other hand, $P_r$ after erase is not as high as $P_r$ after program. This induces weak electron trapping at the top IL ($-Q'_{it}$) and the slight $V_{th}$ shift (2.4 V at a short retention time). Moreover, electron trapping at the top IL is unstable, which deteriorates the retention characteristics after the erase operation.

However, some unclear physics still should be discussed. For instance, the potential questions would be such as "How can the holes be trapped with a large amount at the top SiO₂" and "How can the trapped holes be stable? Where do the holes come from?" We will discuss the questions in the next session.

## III. COMPARISON OF OPERATION MODES

Two different operation modes, hole injection (Fig. 7(a)) and electron de-trapping (Fig. 7(b)), have been suggested to understand the impact of the holes from the channel on the device performance in FEFET [23]. The electron de-trapping mode is carried out by floating the body, suppressing the hole movement at the channel in nFEFET. This leads to a different erase behavior because the hole is not provided enough from the body.

Fig. 8(a) shows $C$-$V$ curves measured by QSCV in MIFIS FEFET. As mentioned, polarization switching peaks during program stay similar, while the erase behaviors by hole injection and electron de-trapping diverge, due to different erase mechanisms. Fig. 8(b) shows that the programmed state by electron de-trapping is much more unstable than by hole injection. Fig. 8(c) also shows that $P_r$ retention after both program and erase by electron de-trapping are degraded.

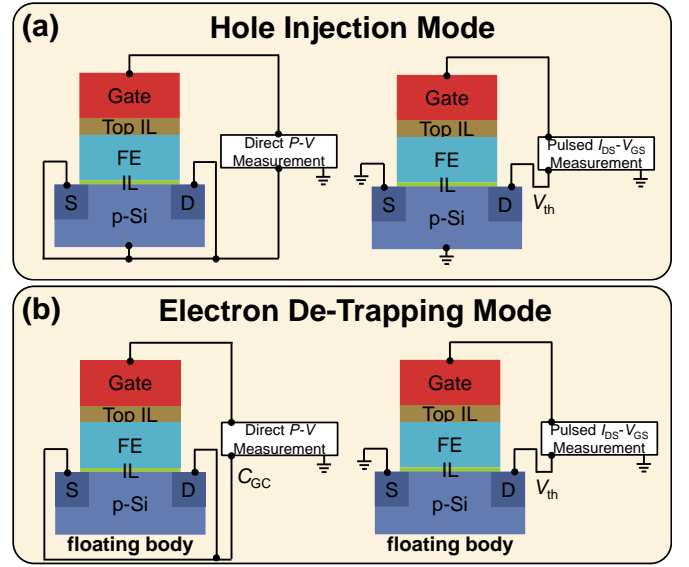

Fig. 7. Schematics of (a) hole injection mode. (b) Schematics of electron de-trapping mode by floating the body [6]. By floating the body, one can suppress hole carrier movements from the body, and hence, hole injection from the channel to the gate stack is suppressed.

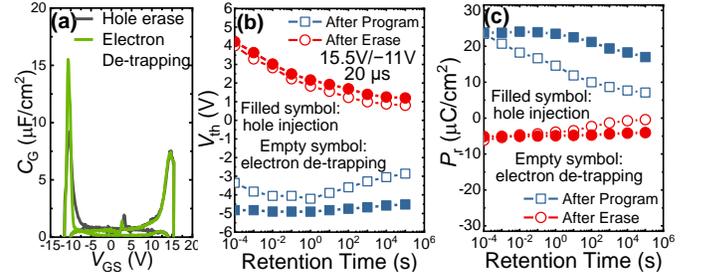

Fig. 8. (a) QSCV (2.5 kHz) comparison between hole injection erase and electron de-trapping erase. It shows the different erase mechanisms (at the negative voltages). (b) $V_{th}$ retention and (c) $P_r$ retention. Data of the hole injection mode are also exhibited for comparison.

The comparison of the operation modes indicates that the hole carrier takes a critical role of enhancing $V_{th}$ and $P_r$ retention after both program and erase. Fig. 9(a) shows proposed energy band diagrams during/after program. During program, electrons are trapped at the gate stack, but the trapped electrons at the top IL move to the gate electrode due to the strong electric field made by high $P_r$ (~$10^{14}$ cm$^{-2}$). On the other hand, holes trapped by erase pulses stay, and additional holes are provided and trapped from the gate electrode [12].

After program, because of the electric field across the top IL and the bottom IL, trapped holes and electrons do not easily de-trap, which provides stable retention characteristics, as also discussed in previous literature [9, 12]. However, in the electron de-trapping mode, holes are not provided enough from the channel during the erase, resulting in the deteriorated retention behavior after program. Note that deteriorated retention is "after program," not only "after erase."

Conversely, during erase (Fig. 9(b)), polarization switching is not as strong as during program. Hence, after erase, electric fields across the top and bottom ILs are weaker compared to after program (Figs. 9(b) and 9(a)). This leads to transient behaviors of trapped electrons at the top IL and unstable

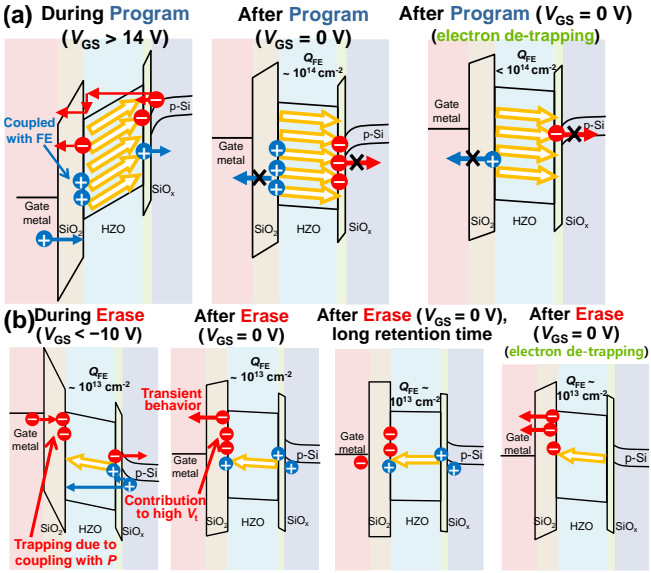

**(a)** During **Program** ($V_{GS}$ > 14 V)    After **Program** ($V_{GS}$ = 0 V)    After **Program** ($V_{GS}$ = 0 V) (electron de-trapping)

**(b)** During **Erase** ($V_{GS}$ < −10 V)    After **Erase** ($V_{GS}$ = 0 V)    After **Erase** ($V_{GS}$ = 0 V, long retention time)    After **Erase** ($V_{GS}$ = 0 V) (electron de-trapping)

Fig. 9. Proposed energy band diagrams during/after (a) program and (b) erase. The orange-colored arrow represents ferroelectric polarization. The amount of switched polarization during/after program is validated by absolute polarization measurement in Figs. 5(c), 6(a) and 8(c). Also, the amount of switched polarization during/after erase is verified by Figs. 5(b), 6(a) and 8(c). Threshold voltage shifts are also measured as Fig. 8(b). Therefore, a novel charge compensation model based on the asymmetric retention characteristics can be established.

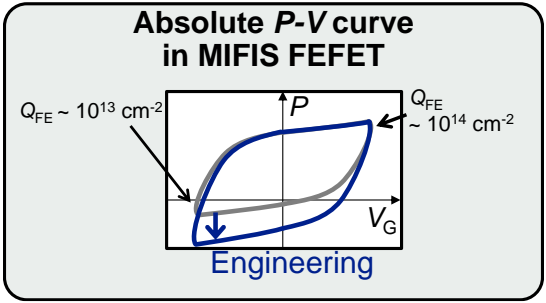

Fig. 10. Proposed additional engineering approach. Increasing polarization switching after erase will contribute to the stable retention characteristics after erase.

Table 1. Benchmarking table of FEFET for flash memory with large MW and feasible gate stack design. Our model is supported by experimental data of $V_{th}$ and absolute polarization, revealing the contribution of asymmetric polarization to the retention behavior for the first time.

| | SK hynix [8] | Samsung [9, 12] | Georgia Tech. [28] | This work |
|---|---|---|---|---|
| **Gate Stack Design** | MIFIS | MIFIS | MFIS | MIFIS |
| **Write Voltage** | 17/−10 V | 16.5/−13 V | 14 V | 15.5/−11 V |
| **Maximum MW** | 10.5 V | 5.5 V | 7.3 V | 9.7 V |
| **Origin of Large MW** | Reduced gate leakage | Charge compensation | Tunneling barrier | Charge compensation by hole (mainly), and electron |
| **Charge Compensation Model** | Based on $V_{th}$ analysis and simulation | Based on $V_{th}$ analysis and simulation | Based on $V_{th}$ analysis and simulation | Based on experimental analysis of $V_{th}$ and absolute polarization; proposing asymmetric retention model after program/erase |

retention characteristics after program.

In this scenario, one might think that hole injection from the channel during erase seems implausible, considering the thicknesses of the oxides and hole-trapping behaviors studied in CTF flash cells [24]. However, the comparison of the two operation modes clearly shows the impact of the hole-injection erase. In particular, the degraded $P_r$ and $V_{th}$ retention after program by the electron de-trapping mode shows that trapped holes at the top IL ($Q'_{it}$) are not enough to screen the polarization, resulting in higher de-polarization fields. Moreover, hole injection depths have been previously studied mostly with the $SiO_2$ dielectric, not $HfO_2$, having lower bandgap energy than $SiO_2$. Therefore, further investigations about the tunneling mechanism of holes from the channel to the top IL, the amounts of trapped carriers, and degradation mechanisms by tunneling and trapping in MIFIS FEFET would be important for industrial implementation.

In summary, holes are provided by both the channel and the gate electrode and trapped at the top IL due to the coupling with ferroelectric polarization. The trapped hole stabilizes both $P_r$ and $V_{th}$ retention after program. On the other hand, because the erase pulses switch polarization only weakly in the MIFIS stack, $P_r$ fails to make enough electric fields for electrons to be stably coupled with at the top IL. This results in the transient behavior of $V_{th}$ retention after erase.

## IV. DEVICE ENGINEERING STRATEGY

The unstable retention phenomenon is similar to the early stage of silicon-oxide-nitride-oxide-silicon (SONOS) technology [24]. SONOS flash also suffered from poor retention characteristics, resulting in the development of bandgap-engineered SONOS [24-25]. Therefore, additional engineering is required to enhance the retention characteristics in MIFIS FEFET as well.

Especially, as we found in this work, the $P_r$ values after program/erase are asymmetric and hence, can be engineered. As found by this work, the poor retention after erase is due to the weak electric field made by low $P_r$ clarified by absolute polarization measurement. Thus, if $P_r$ after erase is increased enough to provide proper electric fields, then the trapped electrons at the top IL would not be easily de-trapped (Fig. 10). This would stabilize the retention after erase, and also would provide the possibility of reducing write voltages by sub-loop operations. As mentioned previously, the bottom IL, $SiO_x$, should be removed or engineered because it has a low dielectric constant (k) hindering polarization switching by erase pulses. Higher-k IL materials (SiON, AlON, etc) [17], or novel channel materials such as SiGe, Ge and oxide semiconductors (Indium Gallium Zinc Oxide (IGZO), etc) might be promising for the approach, as previous studies show IL-free devices with Ge and oxide semiconductor channels [26-27]. These solutions are also extensively discussed and studied in MIFIS FEFETs.

## V. CONCLUSION

We reveal that MIFIS FEFET has unstable retention characteristics especially after erase, and explore the physical origin of the phenomenon. From our investigations, an additional engineering approach has been suggested. Table 1 benchmarks the large MW (9.7 V) and the low write voltages of our device, and emphasizes novelties of our study. Especially, we achieve the lowest write voltage with a large MW, which is important because it contributes to reducing spacer thickness and to further scaling 3D NAND flash. In addition, the experimental observation and understanding of unstable retention characteristics in MIFIS FEFET with the Si channel are first reported by this work. Our model is supported by experimental data of $V_{th}$ and absolute polarization, revealing the contribution of asymmetric polarization to the retention behavior of MIFIS FEFET for the first time.

Because MIFIS FEFET is an "emerging" memory concept, device performance and reliability engineering of retention, endurance, and disturbance are still required. We note again that the unstable retention characteristics of MIFIS FEFET is a similar situation as the beginning phase of the development of SONOS. Additional engineering would solve the reliability issues and finally contribute to realizing FE-NAND or FE-based eFlash. In this context, this work would help researchers to understand the accurate device physics of MIFIS FEFET and to figure out the strategies for further device engineering.